\documentclass[
    ,final            
  ]
  {aipproc}

\layoutstyle{8x11double}

\usepackage{amsmath,amssymb}
\usepackage{mathrsfs}
\usepackage{bm}

\begin{document}

\title{Non-Abelian magnetic monopole dominance for SU(3) Wilson loop average}


\author{Kei-Ichi Kondo}{
  address={Department of Physics,  
Graduate School of Science, 
Chiba University, Chiba 263-8522, Japan}
}

\author{Akihiro Shibata}{
  address={Computing Research Center, High Energy Accelerator Research Organization,  Tsukuba  305-0801, Japan}
}

\author{Toru Shinohara}{
  address={Department of Physics,  
Graduate School of Science, 
Chiba University, Chiba 263-8522, Japan}
}

\author{Seikou Kato}{
  address={Fukui National College of Technology, Sabae 916-8507, Japan}
}

\begin{abstract}
We show that the non-Abelian magnetic monopole defined in a gauge-invariant way in SU(3) Yang-Mills theory gives a dominant contribution to confinement of the fundamental quark, in sharp contrast to the SU(2) case. 

\end{abstract}

\maketitle


\section{Introduction}

The dual superconductor picture proposed long ago  \cite{dualsuper}  is believed to be a promising mechanics for quark confinement. 
For this mechanism to work, however, magnetic monopoles and their condensation are indispensable to cause the dual Meissner effect leading to the linear potential between quark and antiquark, namely,  
  area law of the Wilson loop average.  
The Abelian projection method proposed by 't Hooft \cite{tHooft81} can be used to introduce such magnetic monopoles into the pure Yang-Mills theory even without matter fields. 
Indeed, numerical evidences supporting the dual superconductor picture resulting from such magnetic monopoles have been accumulated since 1990  in pure SU(2) Yang-Mills theory \cite{SY90,SNW94,AS99}.
However, {\it the Abelian projection method explicitly breaks both the local gauge symmetry and the global color symmetry} by partial gauge  fixing from an original non-Abelian gauge group $G=SU(N)$ to the maximal torus subgroup, $H=U(1)^{N-1}$. 
Moreover, the Abelian dominance \cite{SY90} and  magnetic monopole dominance \cite{SNW94} were observed only in a special class of gauges, e.g., the maximally Abelian (MA) gauge and Laplacian Abelian (LA) gauge, realizing the idea of  Abelian projection.

For $G=SU(2)$, we have already succeeded to settle the issue of  gauge (in)dependence by {\it introducing  a gauge-invariant magnetic monopole in a gauge independent way}, based on another method: a non-Abelian Stokes theorem for the Wilson loop operator \cite{DP89,Kondo98b} and a new reformulation of Yang-Mills theory rewritten in terms of new field variables \cite{KMS06,KMS05,Kondo06} and \cite{KKMSSI05,IKKMSS06,SKKMSI07}, elaborating the technique proposed by Cho \cite{Cho80} and Duan and Ge \cite{DG79} independently, and later readdressed by Faddeev and Niemi \cite{FN99}.

For $G=SU(N)$, $N \ge 3$, there are no inevitable reasons why degrees of freedom associated with the maximal torus subgroup should be most dominant for quark confinement.
In this case, the problem is not settled yet. 
In this talk, 
we give a theoretical framework for describing {\it non-Abelian  dual superconductivity} in $D$-dimensional $SU(N)$ Yang-Mills theory, which should be compared with the conventional Abelian $U(1)^{N-1}$ dual superconductivity in $SU(N)$ Yang-Mills theory, hypothesized by  Abelian projection. 
We demonstrate that {\it an effective low-energy description for quarks in the fundamental representation} (abbreviated to rep. hereafter) {\it can be given by a set of non-Abelian restricted field variables} and that {\it non-Abelian $U(N-1)$ magnetic monopoles} in the sense of Goddard--Nuyts--Olive--Weinberg \cite{nAmm} {\it are the most dominant topological configurations for quark confinement} as conjectured in \cite{KT99,Kondo99Lattice99}.


\section{Wilson loop and gauge-inv. magnetic monopole}

A version of a non-Abelian Stokes theorem (NAST) for the Wilson loop operator originally invented by Diakonov and Petrov \cite{DP89} for $G=SU(2)$ was proved to hold \cite{Kondo98b} and was extended to $G=SU(N)$ \cite{KT99,Kondo08} in a unified way \cite{Kondo08} as a path-integral rep. by making use of a coherent state for the Lie group.  
For the Lie algebra $su(N)$-valued Yang-Mills field $\mathscr{A}_\mu(x)=\mathscr{A}_\mu^A(x) T_A$ with $su(N)$ generators $T_A$ ($A=1, \cdots, N^2-1$), 
the NAST enables one to rewrite a non-Abelian Wilson loop operator \begin{align}
  W_C[\mathscr{A}]  
:=& {\rm tr} \left[ \mathscr{P} \exp \left\{ ig_{\rm YM} \oint_{C} dx^\mu \mathscr{A}_\mu(x) \right\} \right]/{\rm tr}({\bf 1})  
 ,
\end{align}
into the surface-integral form:
\begin{equation}
 W_C[\mathscr{A}] = \int d\mu_{\Sigma}(g) \exp \left[  ig_{\rm YM} \int_{\Sigma: \partial \Sigma=C} F \right] ,
\end{equation}
where $d\mu_{\Sigma}(g):=\prod_{x \in \Sigma} d\mu(g_{x})$, with an invariant measure $d\mu$ on $G$  normalized as $\int d\mu(g_{x})=1$, $g_{x}$ is an element of a gauge group $G$ (more precisely, rep. $D_R(g_{x})$ of  $G$),
the two-form $F:=dA=\frac12 F_{\mu\nu}(x) dx^\mu \wedge dx^\nu$ is defined from the one-form $A := A_\mu(x) dx^\mu$, 
$
A_\mu(x) = {\rm tr}\{ \rho[ g_{x}^\dagger \mathscr{A}_\mu(x) g_{x} + ig_{\rm YM}^{-1} g_{x}^\dagger \partial_\mu g_{x} ] \} ,  
$ 
by
\begin{align}
  F_{\mu\nu}(x) &=  \sqrt{2(N-1)/N} [\mathscr{G}_{\mu\nu} (x)   
+ ig_{\rm YM}^{-1} {\rm tr} \{ \rho g_{x}^\dagger [\partial_\mu, \partial_\nu] g_{x} \} ],
\end{align}
with the field strength $\mathscr{G}_{\mu\nu}$ defined by
\begin{align}
 \mathscr{G}_{\mu\nu} (x) 
  &:=   \partial_\mu {\rm tr} \{ \mathbf{n}(x) \mathscr{A}_\nu(x) \} - \partial_\nu {\rm tr} \{ \mathbf{n}(x) \mathscr{A}_\mu(x) \} 
\nonumber\\& 
+ \frac{2(N-1)}{N} ig_{\rm YM}^{-1} {\rm tr} \{ \mathbf{n}(x) [\partial_\mu \mathbf{n}(x), \partial_\nu \mathbf{n}(x) ] \} 
 ,
\end{align}
and a normalized traceless field $\mathbf{n}(x)$ called the color field  
\begin{equation}
 \mathbf{n}(x) :=  \sqrt{N/[2(N-1)]} g_{x} \left[ \rho - \bm{1}/{\rm tr}(\bm{1}) \right] g_{x}^\dagger .
\end{equation}
Here $\rho$ is defined as $\rho :=  | \Lambda \rangle \langle \Lambda |$ using a reference state (highest or lowest  weight state of the rep.) $| \Lambda  \rangle$ making a rep. of the Wilson loop we consider. 
Note that ${\rm tr}(\rho) = \langle \Lambda | \Lambda \rangle = 1$ follows from the normalization of $| \Lambda \rangle$. 

Finally, the Wilson loop operator in the fundamental rep. of $SU(N)$ reads \cite{Kondo08}
\begin{align}
& W_C[\mathscr{A}] 
=  \int  d\mu_{\Sigma}(g)  \exp \left\{  ig_{\rm YM} (k, \Xi_{\Sigma}) + ig_{\rm YM} (j, N_{\Sigma}) \right\} ,
\label{NAST-SUN}
\nonumber\\
& k:=   \delta *f = *df, \quad j:=  \delta f , 
\quad
f:=  \sqrt{2(N-1)/N}  \mathscr{G}  ,
\nonumber\\
& \Xi_{\Sigma} :=  * d\Theta_{\Sigma} \Delta^{-1} = \delta *\Theta_{\Sigma} \Delta^{-1} , \
 N_{\Sigma} := \delta \Theta_{\Sigma} \Delta^{-1} ,
\end{align}
where two conserved currents,   ``magnetic-monopole current''   $k$ and  ``electric current'' $j$, are introduced, 
$\Delta:=d\delta+\delta d$ is the $D$-dimensional Laplacian, and $\Theta$ is an antisymmetric tensor of rank two  called the vorticity tensor:
$
 \Theta^{\mu\nu}_{\Sigma}(x) 
:=   \int_{\Sigma}  d^2S^{\mu\nu}(x(\sigma)) \delta^D(x-x(\sigma))  ,
$
which  has the support on the surface $\Sigma$ (with the surface element $dS^{\mu\nu}(x(\sigma))$) whose boundary is the loop $C$.
Incidentally, the last part $ig_{\rm YM}^{-1} {\rm tr} \{ \rho g_{x}^\dagger [\partial_\mu, \partial_\nu] g_{x} \}$ in $F$ corresponds to the Dirac string \cite{Kondo97,Kondo98a}, which is not gauge invariant and does not contribute to the Wilson loop in the end. 

For $SU(3)$ in the fundamental rep., the lowest-weight state $\langle \Lambda |=(0,0,1)$ leads to 
\begin{equation}
 \mathbf{n}(x) = g_{x} (\lambda_8/2) g_{x}^\dagger \in SU(3)/[SU(2) \times U(1)] \simeq CP^2 ,
\end{equation}
with the Gell-Mann matrix $\lambda_8:={\rm diag.}(1,1,-2)/\sqrt{3}$, 
while for $SU(2)$, $\langle \Lambda |=(0,1)$ yields
\begin{equation}
 \mathbf{n}(x) = g_{x} (\sigma_3/2) g_{x}^\dagger \in SU(2)/U(1) \simeq S^2 \simeq CP^1 ,
\end{equation}
with the Pauli matrix $\sigma_3:={\rm diag.}(1,-1)$.
The existence of magnetic monopole can be seen by a nontrivial Homotopy class of the map $\mathbf{n}$ from $S^2$ to the target space of the color field $\mathbf{n}$ \cite{KT99}:  
For $SU(3)$, 
\begin{align}
 & \pi_2(SU(3)/[SU(2) \times U(1)])=\pi_1(SU(2) \times U(1)) 
\nonumber\\&
=\pi_1(U(1))=\mathbb{Z} ,
\end{align}
while for $SU(2)$
\begin{equation}
 \pi_2(SU(2)/U(1))=\pi_1(U(1))=\mathbb{Z} .
\end{equation}
For $SU(3)$, the magnetic charge of the non-Abelian magnetic monopole obeys the quantization condition \cite{Kondo08}:
\begin{equation}
 Q_m := \int d^3x k^0 = 2\pi \sqrt{3} g_{\rm YM}^{-1} n , \ n \in \mathbb{Z} .
\end{equation}
The NAST shows that {\it the $SU(3)$ Wilson loop operator in the fundamental rep. detects the inherent $U(2)$ magnetic monopole which is  $SU(3)$ gauge invariant}. 
The rep. can be classified by its {\it stability group} $\tilde H$ of $G$ \cite{KT99,Kondo08}.  
For the fundamental rep. of $SU(3)$, the stability group  is $U(2)$.  
Therefore, the non-Abelian $U(2) \simeq SU(2)  \times U(1)$ magnetic monopole follows from  $\tilde H=SU(2)_{1,2,3} \times U(1)_{8}$, while the Abelian $U(1) \times U(1)$ magnetic monopole comes from $\tilde H=U(1)_{3} \times U(1)_{8}$. 
The adjoint rep. belongs to the latter case. 
The former case occurs only when the weight vector of the rep. is orthogonal to some of root vectors.  
The fundamental rep. is indeed  this case. 
For $SU(2)$, such a difference does not exist and $U(1)$ magnetic monopoles appear, since $\tilde H$ is always $U(1)$ for any rep.. 
For $SU(3)$, our result is different from Abelian projection: two independent $U(1)$ magnetic monopoles appear for any rep., since 
\begin{align}
 & \pi_2(SU(3)/U(1) \times U(1))=\pi_1(U(1) \times U(1)) =\mathbb{Z}^2 .
\end{align}

\section{Numerical simulations}

The SU(3) Yang-Mills theory can be reformulated in the continuum and on a lattice using new variables. 
 For  $SU(3)$, two options  are possible, maximal for $\tilde H=U(1)^2$ \cite{Cho80c,FN99a} and minimal for $\tilde H=U(2)$ \cite{KSM08}. 
In our reformulation,  all the new variables  $\mathscr{C}_\mu$, $\mathscr{X}_\mu$ and $\mathbf{n}$  are obtained 
from   $\mathscr{A}_\mu$: 
\begin{equation}
 \mathscr{A}_\mu^A  \Longrightarrow (\mathbf{n}^\beta, \mathscr{C}_\nu^k,  \mathscr{X}_\nu^b)  
 ,
\end{equation} 
once the color field $\mathbf{n}$ is determined by solving the reduction condition:
\begin{equation}
\bm\chi[\mathscr{A},\mathbf{n}]
 :=[ \mathbf{n} ,  D^\mu[\mathscr{A}]D_\mu[\mathscr{A}]\mathbf{n} ]
 = 0 
  ,
\label{eq:diff-red}
\end{equation}

On a four-dimensional Euclidean lattice, gauge field configurations  $\{ U_{x,\mu} \}$ are generated by using the standard Wilson action and pseudo heat-bath method.  
For a given $\{ U_{x,\mu} \}$,   color field $\{ \bm{n}_{x} \}$ are determined by imposing a lattice version of  reduction condition.  Then new variables are introduced by using the lattice version of  change of variables \cite{lattice-f}. 

\begin{figure}[h]
\includegraphics[height=4.0cm,width=7.0cm]{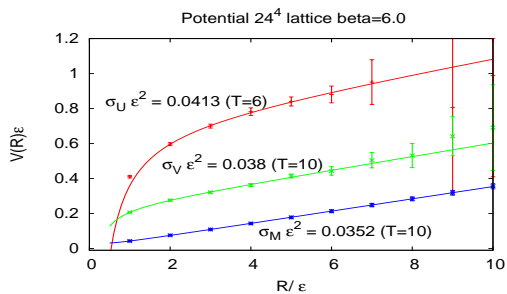}
\vspace{-0.6cm}
\caption{$SU(3)$ quark-antiquark potential: (from above to below) 
 full potential $V_f(r)$,  restricted part $V_a(r)$  and  magnetic--monopole part  $V_m(r)$  at $\beta=6.0$ on $24^4$ ($\epsilon$: lattice spacing).}
\label{fig:quark-potential}
\end{figure}

Fig. \ref{fig:quark-potential} shows the full $SU(3)$ quark-antiquark potential $V(r)$ obtained from the $SU(3)$ Wilson loop average $\langle W_C[\mathscr{A}] \rangle$, the restricted part $V_a(r)$ from the  $\mathscr{V}$ Wilson loop average $\langle W_C[\mathscr{V}] \rangle$, and  magnetic--monopole part  $V_m(r)$ from    $\langle e^{  ig_{\rm YM} (k, \Xi_{\Sigma})  }  \rangle$. They are gauge invariant quantities by construction.  These results exhibit infrared $\mathscr{V}$ dominance in the string tension (85--90\%) and  non-Abelian $U(2)$ magnetic monopole dominance in the string tension  (75\%) in the gauge independent way.


\section{Conclusion}

We have shown: 
(i) The $SU(N)$ Wilson loop operator can be rewritten in terms of a pair of gauge-invariant magnetic-monopole current  $k$ ($(D-3)$-form) and the associated geometric object defined from the Wilson surface $\Sigma$ bounding the Wilson loop $C$, and another pair of an electric current $j$ (one-form  independently of $D$) and the associated topological object, which follows from a non-Abelian Stokes theorem for the Wilson loop operator \cite{Kondo08}.   
(ii) The $SU(N)$ Yang-Mills theory can be reformulated in terms of new field variables obtained by change of variables from the original Yang-Mills gauge field $\mathscr{A}_\mu^A(x)$ \cite{KSM08}, so that it gives an optimal description for the non-Abelian magnetic monopole defined from the $SU(N)$ Wilson loop operator in the fundamental rep. of quarks.
(iii) A lattice version of the reformulated  Yang-Mills theory can be constructed \cite{lattice-f}.
Numerical simulations of the  lattice $SU(3)$ Yang-Mills theory give numerical evidences that the restricted field variables become dominant in the infrared for correlation functions and the string tension ({\it infrared restricted non-Abelian dominance}) and that the $U(2)$ magnetic monopole gives a most dominant contribution to the string tension obtained from  $SU(3)$ Wilson loop average ({\it non-Abelian magnetic monopole dominance}). 
See \cite{KSSK10} for more informations.




\begin{theacknowledgments}
This work is  supported by Grant-in-Aid for Scientific Research (C) 21540256 from Japan Society for the Promotion of Science
(JSPS).

\end{theacknowledgments}



\bibliographystyle{aipproc}   





\end{document}